# Performance of Cognitive Radio Systems over $\kappa - \mu$ Shadowed with Integer $\mu$ and Fisher-Snedecor $\mathcal{F}$ Fading Channels

Hussien Al-Hmood
Electrical and Electronics Engineering Department, Thi-Qar University
Thi-Qar, IRAQ
Emails: {hussien.al-hmood}@eng.utq.edu.iq, @brunel.ac.uk

*Abstract*— In this paper, we analyze the performance of cognitive radio (CR) systems over different composite generalized multipath /shadowed fading scenarios. The $\kappa - \mu$ shadowed and Fisher-Snedecor $\mathcal{F}$ fading channels which are proposed as a simple and high accurate distributions in comparison with generalized-K ($K_G$) and Nakagami-$m$ shadowed conditions are used in this analysis. For the $\kappa - \mu$ shadowed, a novel simple exact closed-form analytic expression for the probability density function (PDF) and the cumulative distribution function (CDF) are introduced by assuming the fading parameters are integer numbers. To this end, the detection performance metrics, namely, the average detection probability and the average area under the receiver operating characteristics curve (AUC) which are used in the analysis of energy detection and the effective rate and the effective rate are derived. To validate the results of this work, comparisons between the simulated and numerical results as well as with various conventional channel models and scenarios are given.

## I. INTRODUCTION

To represent the impact of multipath of wireless channels, various formats of distributions have been employed such as Rayleigh, Nakagami-$m$, and Nakagami-$n$ [1]. In addition to that, the wireless channels may suffer from the shadowing effects. Hence, different composite multipath/shadowing fading conditions have been proposed by several works to model the statistics of the signal-to-noise ratio (SNR) [2]-[4].

Recently, different generalized composite distributions such as gamma shadowed $\kappa - \mu$, $\eta - \mu$, $\alpha - \mu$, and $\alpha - \kappa - \mu$, have been suggested to model various communication scenarios [5]-[8]. This is because they can provide close results to the practical measurements in comparison with the conventional distributions. Moreover, most of the well-known distributions can be deduced from the aforementioned generalized formats. Accordingly, these distributions have been widely utilized in the analysis of communication systems. For example, in [5], the probability density function (PDF) and cumulative distribution function (CDF) of the SNR in $\kappa - \mu$ shadowed ($\kappa - \mu$/Nakagami-$m$) are derived with applications to communications systems. In addition, this distribution is used to analyze the behavior of the energy detection (ED) that is utilized to provide spectrum sensing in cognitive radio networks and ultra-wide-band systems [6], [9]-[12]. However, the results are either approximated [6] or included an infinite series [9]-[12]. In [13] and [14], the effective rate analysis over $\kappa - \mu$ shadowed and unified fading models are, respectively, investigated. But, the expression is included a multivariate Meijer's G-function that is not yet presented in most widely used mathematical software packages such as MATLAB and MATHEMATICA. This is because the derived PDFs in [5] and [15] would lead to more complicated mathematical expressions when they are employed in the study the performance of the communication systems.

In this paper, the PDF of the SNR over $\kappa - \mu$ shadowed fading condition is investigated in simple closed-form expressions. In contrast to previous work in [5] and [15], the fading parameters $\mu$ and $m$ are assumed to be integer numbers. Accordingly, the derived results are simple where they expressed in terms of power and exponential functions. Moreover, no work has been done to study the behavior of ED and effective rate over $\kappa - \mu$ shadowed fading by using integer numbers for both $\mu$ and $m$. To this effect, the detection performance metrics, such as the average probability of detection, and the average area under the receiver operating characteristics curve (AUC) and the effective rate, are given in simple exact closed-from expressions.

More recent, the Fisher-Snedecor $\mathcal{F}$ fading channel has been proposed as a composite of Nakagami-$m$/inverse Nakagami-$m$ to model device-to-device (D2D) fading channels at 5.8 GHz in both indoor and outdoor environments [16]. Accordingly, the expression for the aforementioned performance metrics over Fisher-Snedecor $\mathcal{F}$ fading channel. This distribution can be used for both LoS and NLoS communication scenarios with results closer to the empirical data than the $K_G$ fading channel. To the best of our knowledge, the Fisher-Snedecor $\mathcal{F}$ fading scenario has not been yet used in the literature in the analysis of ED and the effective rate of the cognitive radio (CR) systems.

The rest of this work is structured as follows. The PDF of $\kappa - \mu$ and Fisher-Snedecor $\mathcal{F}$ fading channels are given in Section II and Section III, respectively. Section IV provides the performance metrics that are used in the analysis of ED. The effective rate is analyzed in Section V. The numerical and simulation results are given in Section VI. Section VII concludes the presented ideas.

## II. SIMPLE CLOSED-FORM PDF OF $\kappa - \mu$ SHADOWED FADING SCENARIOS

The moment generating function (MGF) of the SNR, $\gamma$, over $\kappa - \mu$ shadowed fading scenario is expressed as [5, eq. (5)]

$$\mathcal{M}_\gamma(s) = \frac{1}{\left(1 - \frac{s}{\theta_1}\right)^{\mu-m} \left(1 - \frac{s}{\theta_2}\right)^m} \quad (1)$$

where $\theta_1 = \frac{\mu(1+\kappa)}{\bar{\gamma}}$, $\theta_2 = \frac{m}{(\mu\kappa+m)}\theta_1$, $\bar{\gamma}$ is the average SNR, $\kappa$ stands for the ratio between the total powers of the dominant components and scattered waves, $\mu$ indicates the number of the multipath clusters, and $m$ represents the shadowing severity index.

It can be noted that (1) can be written as $\mathcal{M}_\gamma(s) = \mathcal{M}_{\gamma_1}(s) \cdot \mathcal{M}_{\gamma_2}(s)$ with

$$\mathcal{M}_{\gamma_1}(s) = \frac{1}{\left(1 - \frac{s}{\theta_1}\right)^{\mu-m}}$$

and

$$\mathcal{M}_{\gamma_2}(s) = \frac{1}{\left(1 - \frac{s}{\theta_2}\right)^m} \quad (2)$$

The PDF of $\mathcal{M}_{\gamma_1}(s)$ and $\mathcal{M}_{\gamma_2}(s)$ can be computed by using the inverse Laplace transform, i.e., $f_\gamma(\gamma) = \mathcal{L}^{-1}[\mathcal{M}_\gamma(s); \gamma]$, as follows

$$f_{\gamma_1}(\gamma) = \frac{\theta_1^{\mu-m}}{\Gamma(\mu-m)} \gamma^{\mu-m-1} e^{-\theta_1 \gamma}$$

and

$$f_{\gamma_2}(\gamma) = \frac{\theta_2^m}{\Gamma(m)} \gamma^{m-1} e^{-\theta_2 \gamma} \quad (3)$$

where $\Gamma(c) = \int_0^\infty y^{c-1} e^{-y} dy$ is the Gamma function.

To find the PDF of the $\kappa - \mu$ shadowed condition, the convolution between $f_{\gamma_1}(\gamma)$ and $f_{\gamma_2}(\gamma)$ is carried out as follows

$$f_\gamma(\gamma) = f_{\gamma_1}(\gamma) \otimes f_{\gamma_2}(\gamma) = \int_0^\gamma f_{\gamma_1}(x) f_{\gamma_2}(\gamma - x) \, dx \quad (4)$$

Substituting (3) in (5), this yields

$$f_\gamma(\gamma) = \frac{\theta_1^{\mu-m} \theta_2^m}{\Gamma(\mu-m)\Gamma(m)} e^{-\theta_2 \gamma}$$
$$\times \int_0^\gamma x^{\mu-m-1} (\gamma - x)^{m-1} e^{-(\theta_1-\theta_2)x} dx \quad (5)$$

Assuming $m \in \mathbb{Z}^+$ and using $(a-b)^c = \sum_{i=0}^c \binom{c}{i} a^{c-i}(-b)^i$ [17, eq. (1.111), p. 25] and [17, eq. (3.351.1), p. 340], a simple closed-form expression of the PDF of $\kappa - \mu$ shadowed is deduced

$$f_\gamma(\gamma) = \frac{\theta_1^{\mu-m} \theta_2^m}{\Gamma(\mu-m)\Gamma(m)} e^{-\theta_2 \gamma} \sum_{i=0}^{m-1} \binom{m-1}{i} (-1)^i \gamma^{m-i-1}$$
$$\times \frac{G(\mu-m+i, (\theta_1-\theta_2)\gamma)}{(\theta_1-\theta_2)^{\mu-m+i}} \quad (5)$$

where $\binom{a}{b} \triangleq \frac{a!}{(a-b)!}$ is the binomial coefficients and $G(z, y) = \int_0^y x^{z-1} e^{-x} dx$ is the lower incomplete Gamma function.

When $\mu \in \mathbb{Z}^+$, we can use the identity $G(z, y) = \Gamma(z)\left[1 - e^{-y} \sum_{k=0}^{z-1} \frac{y^k}{k!}\right]$ [17, eq. (8.352.1), p. 899] in (5). Hence, we have

$$f_\gamma(\gamma) = \frac{\theta_1^{\mu-m} \theta_2^m}{\Gamma(m)} e^{-\theta_2 \gamma} \sum_{i=0}^{m-1} \binom{m-1}{i} (-1)^i \gamma^{m-i-1} (\mu-m)_i$$
$$\times \left[1 - e^{-(\theta_1-\theta_2)\gamma} \sum_{k=0}^{\mu-m+i-1} \frac{((\theta_1-\theta_2)\gamma)^k}{k!}\right] \quad (6)$$

where $(.)_i$ is the Pochhammer symbol.

## III. THE PDF OF FISHER-SNEDECOR $\mathcal{F}$ FADING SCENARIO

The PDF of the SNR, $\gamma$, using Fisher-Snedecor $\mathcal{F}$ distribution is given as [16, eq. (5)]

$$f_\gamma(\gamma) = \frac{\Omega^m}{B(m, m_s)} (1 + \Omega\gamma)^{-(m+m_s)} \gamma^{m-1} \quad (7)$$

where $\Omega = \frac{m}{m_s \bar{\gamma}}$, $m$, $m_s$, $\bar{\gamma}$ and $B(c_1, c_2) = \Gamma(c_1)\Gamma(c_2)/\Gamma(c_1 + c_2)$ are the number of multipath clusters, the shape parameter, the average SNR and the beta function, respectively.

## IV. Performance Analysis of Energy Detection Based Spectrum Sensing of Cognitive Radio User

### A. Average Probability of Detection

The instantaneous detection, $P_d(\gamma, \lambda)$, and false alarm, $P_f(\lambda)$, probabilities for the ED model that is employed in this work are, respectively, given by [18]

$$P_d(\gamma, \lambda) = Q_u(\sqrt{2\gamma}, \sqrt{\lambda}) \tag{8}$$

and

$$P_f(\lambda) = \frac{\Gamma(u, \lambda/2)}{\Gamma(u)} \tag{9}$$

where $u$ is the time-bandwidth product, $Q_u(.,.)$ is the $u$th order generalized Marcum-Q function, and $\Gamma(z, y) = \int_y^\infty x^{z-1} e^{-x} dx$ is the upper incomplete Gamma function.

The average probability of detection, $\bar{P}_d(\lambda)$, can be evaluated by [6, eq. (10)]

$$\bar{P}_d(\lambda) = \int_0^\infty P_d(\gamma, \lambda) f_\gamma(\gamma) d\gamma \tag{10}$$

*1) $\kappa - \mu$ shadowed fading channel:* Plugging (6) and (8) in (10), this yields

$$\bar{P}_d(\lambda) = \frac{\theta_1^{\mu-m} \theta_2^m}{\Gamma(m)} \sum_{i=0}^{m-1} \binom{m-1}{i} (-1)^i (\mu - m)_i$$
$$\times \left[ \int_0^\infty \gamma^{m-i-1} e^{-\theta_2 \gamma} Q_u(\sqrt{2\gamma}, \sqrt{\lambda}) d\gamma - \sum_{k=0}^{\mu-m+i-1} \frac{(\theta_1 - \theta_2)^k}{k!} \right.$$
$$\left. \int_0^\infty \gamma^{m+k-i-1} e^{-\theta_1 \gamma} Q_u(\sqrt{2\gamma}, \sqrt{\lambda}) d\gamma \right] \tag{11}$$

With the help of [19, eq. (3)], both integrals in (11) can be calculated in exact closed-form as follows

$$\bar{P}_d(\lambda) = \frac{\theta_1^{\mu-m} \theta_2^m}{\Gamma(m)} \sum_{i=0}^{m-1} \binom{m-1}{i} (-1)^i \Gamma(m-i)(\mu - m)_i$$
$$\left\{ \frac{1}{\theta_2^{m-i}} \left[ P_f(\lambda) + \sum_{j=0}^{m-i-1} \mathcal{B}_2(j) \right] - \sum_{k=0}^{\mu-m+i-1} \binom{m-i}{k} \frac{(\theta_1 - \theta_2)^k}{\theta_1^{m-i+k}} \right.$$
$$\left. \left[ P_f(\lambda) + \sum_{j=0}^{m-i+k-1} \mathcal{B}_1(j) \right] \right\} \tag{12}$$

where $\mathcal{B}_t(j) = \frac{\lambda^u e^{-\frac{\lambda}{2}} \theta_t^j}{2^u u! (1+\theta_t)^{j+1}} \, _1F_1\left(j+1; u+1; \frac{\lambda}{2(1+\theta_t)}\right)$ with $t \in \{1, 2\}$ and $_1F_1(.;.;.)$ is confluent hypergeometric function defined [17, eq. (9.14.1), p. 1010]. In contrast to [6], [9]-[12] in which the $\bar{P}_d(\lambda)$ over $\kappa - \mu$ shadowed fading channel is either approximated or included an infinite series, (12) is given in exact closed-form expression.

*2) Fisher-Snedecor $\mathcal{F}$ Fading Channel:* Inserting (7) and (8) in (10) with the aid of [20, eq. (2.9), p. 23], we have

$$\bar{P}_d(\lambda) = \frac{\Omega^m}{B(m, m_s)} \sum_{j=0}^\infty \frac{\Gamma(j + u, \lambda/2)}{\Gamma(j + u) j!}$$
$$\times \int_0^\infty \gamma^{j+m-1} (1 + \Omega\gamma)^{-(m+m_s)} e^{-\gamma} d\gamma \tag{13}$$

Employing [20, eq. (2.9)] to evaluate the integration in (13) and doing some simple straightforward mathematical operations, the result is

$$\bar{P}_d(\lambda) = \frac{1}{B(m, m_s)} \sum_{j=0}^\infty \frac{\Gamma(j + u, \lambda/2) \Gamma(j + m)}{\Omega^j \Gamma(j + u) j!}$$
$$\times U\left(j + m; j - m_s + 1; \frac{1}{\Omega}\right) \tag{14}$$

where $U(.;.;.)$ is the confluent Tricomi hypergeometric function of the second kind.

One can observe that (14) is expressed in terms of an infinite series. Thus, a convergence by a limited number of terms, $S$, with truncation error $|E_S|$ can be performed. Accordingly, by utilizing the identity $\Gamma(x, y) = \Gamma(x) - G(x, y)$ [17, eq. (8.356.1), p. 900], this yields

$$|E_S| = \sum_{j=S}^\infty \frac{\Gamma(j + m)}{\Omega^j j!} U\left(j + m; j - m_s + 1; \frac{1}{\Omega}\right)$$
$$- \sum_{j=S}^\infty \frac{G(j + u, \lambda/2) \Gamma(j + m)}{\Omega^j \Gamma(j + u) j!} U\left(j + m; j - m_s + 1; \frac{1}{\Omega}\right) \tag{15}$$

It can be noticed that both $U(.;.;.)$ and $G(.,.)$ in (15) are decreasing with $j$. Consequently, after following the same procedure in [19] and performing some mathematical manipulations, the result is

$$|E_S| \leq \frac{\Gamma(S + m)}{\Omega^S S!} U\left(S + m; S - m_s + 1; \frac{1}{\Omega}\right)$$
$$\left[ \sum_{j=0}^\infty \frac{(S + m)_j (1)_j}{(S + 1)_j j!} \left(\frac{1}{\Omega}\right)^j - \frac{G(S + u, \lambda/2)}{\Gamma(S + u)} \right.$$
$$\left. \sum_{j=0}^\infty \frac{(S + m)_j (1)_j}{(S + 1)_j (S + u)_j j!} \left(\frac{1}{\Omega}\right)^j \right] \tag{16}$$

To represent the infinite series of (16) in closed-from, we can employ the confluent hypergeometric function defined by [17, eq. (9.14.1), p. 1010]

$$_aF_b(x_1, \ldots, x_a; y_1, \ldots, y_b; z) = \sum_{l=0}^{\infty} \frac{(x_1)_l \ldots (x_a)_l}{(y_1)_l \ldots (y_b)_l} \frac{z^l}{l!} \quad (17)$$

Hence, the following closed-form expression is yielded

$$|E_S| \leq \frac{\Gamma(S+m)}{\Omega^S S!} U\left(S+m; S-m_s+1; \frac{1}{\Omega}\right)$$
$$\left[{}_2F_1\left(S+m, 1; S+1; \frac{1}{\Omega}\right) - \frac{G(S+u, \lambda/2)}{\Gamma(S+u)}\right.$$
$$\left.\sum_{j=0}^{\infty} {}_2F_2\left(S+m, 1; S+1, S+u; \frac{1}{\Omega}\right)\right] \quad (18)$$

### B. Average Area under the ROC (AUC)

The average AUC, $\bar{A}$, can be computed by [12, eq. (4)]

$$\bar{A} = \int_0^{\infty} A(\gamma) f_\gamma(\gamma) d\gamma \quad (19)$$

For the ED model that is given in (8) and (9), $A(\gamma)$ is given as [12, eq. (1)]

$$A(\gamma) = 1 - \sum_{l=0}^{u-1} \sum_{i=0}^{l} \binom{l+u-1}{l-i} \left(\frac{1}{2}\right)^{l+i+u} \frac{\gamma^i e^{-\left(\frac{\gamma}{2}\right)}}{i!} \quad (20)$$

*1) $\kappa - \mu$ shadowed fading channel:* Plugging (6) and (20) in (19), the following integral should be evaluated [17, eq. (3.381.3), p. 346]

$$I = \int_0^{\infty} x^{a-1} e^{-bx} dx = \frac{\Gamma(a)}{b^a} \quad (21)$$

Using (21) and $\int_0^{\infty} f_\gamma(\gamma) d\gamma \triangleq 1$ with some mathematical simplifications, this yields

$$A(\gamma) = 1 - \sum_{l=0}^{u-1} \sum_{i=0}^{l} \binom{l+u-1}{l-i} \left(\frac{1}{2}\right)^{l+i+u} \frac{\theta_1^{\mu-m} \theta_2^m}{\Gamma(m) i!}$$
$$\sum_{i=0}^{m-1} \binom{m-1}{i} (-1)^i \gamma^{m-i-1} (\mu-m)_i \left[\frac{\Gamma(n+m-i)}{\left(\frac{1}{2}+\theta_2\right)^{n+m-i}}\right.$$

$$\left. - \sum_{k=0}^{\mu-m+i-1} \frac{(\theta_1 - \theta_2)^k}{k!} \frac{\Gamma(n+m+k-i)}{\left(\frac{1}{2}+\theta_2\right)^{n+m+k-i}}\right] \quad (22)$$

*2) Fisher-Snedecor $\mathcal{F}$ Fading Channel:* Substituting (7) and (20) in (19) and using $\int_0^{\infty} f_\gamma(\gamma) d\gamma \triangleq 1$, we have

$$\bar{A} = 1 - \frac{\Omega^m}{B(m, m_s)} \sum_{l=0}^{u-1} \sum_{i=0}^{l} \binom{l+u-1}{l-i} \left(\frac{1}{2}\right)^{l+i+u} \frac{1}{i!}$$
$$\times \int_0^{\infty} \gamma^{i+m-1} (1+\Omega\gamma)^{-(m+m_s)} e^{-\frac{\gamma}{2}} d\gamma \quad (23)$$

Invoking [20, eq. (2.9)], (23) can be evaluated in closed-form as follows

$$\bar{A} = 1 - \frac{1}{B(m, m_s)} \sum_{l=0}^{u-1} \sum_{i=0}^{l} \binom{l+u-1}{l-i} \left(\frac{1}{2}\right)^{l+i+u} \frac{\Gamma(i+m)}{\Omega^i i!}$$
$$\times U\left(i+m; i-m_s+1; \frac{1}{2\Omega}\right) \quad (24)$$

## V. EFFECTIVE RATE ANALYSIS OF COGNITIVE RADIO USER

The effective rate, $\mathcal{R}$, can be evaluated by [13, eq. (1)]

$$\mathcal{R} = -\frac{1}{A} \log_2 \left(\int_0^{\infty} (1+\gamma)^{-A} f_\gamma(\gamma) d\gamma\right) \quad (25)$$

where $A \triangleq \Theta TB/\ln 2$, $\Theta$, $T$, and $B$ denote the delay exponent, block duration, and bandwidth of the system, respectively.

*1) $\kappa - \mu$ shadowed fading channel:* Inserting (6) in (25) and utilizing [20, eq. (2.9)] to compute the integral, the result is

$$\mathcal{R} = -\frac{1}{A} \log_2 \left(\frac{\theta_1^{\mu-m} \theta_2^m}{\Gamma(m)} \sum_{i=0}^{m-1} \binom{m-1}{i} (-1)^i (\mu-m)_i\right.$$
$$\times [\Gamma(m-i) U(m-i; m-i-A+1; \theta_2)$$
$$- \sum_{k=0}^{\mu-m+i-1} \frac{(\theta_1-\theta_2)^k}{k!} \Gamma(m-i+k)$$
$$\left. \times U(m-i+k; m-i+k-A+1; \theta_1)]\right) \quad (26)$$

*2) Fisher-Snedecor $\mathcal{F}$ Fading Channel:* Plugging (7) in (25), this yields

$$\mathcal{R} = -\frac{1}{A} \log_2 \left(\frac{\Omega^m}{B(m, m_s)}\right.$$
$$\left.\int_0^{\infty} \gamma^{m-1} (1+\Omega\gamma)^{-(m+m_s)} (1+\gamma)^{-A} d\gamma\right) \quad (27)$$

With the aid of [17, eq. (3.197.1), p. 317], (27) can be written in exact expression as follows

$$\mathcal{R} = -\frac{1}{A}\log_2\left(\frac{\Omega^m}{B(m,m_s)}\,B(m,m_s+A)\right.$$
$$\left.\times\,_2F_1\left(m+m_s,m;m+m_s+A;1-\frac{m}{m_s\bar{\gamma}}\right)\right) \qquad (28)$$

## VI. ANALYTICAL RESULTS

To validate the derived expressions, this section provides the numerical and simulated results which are generated by Monte Carlo for $10^6$ iterations. In all figures, the solid lines stand for the simulated results whereas the marks represent the numerical counterparts. The parameter, $S$, that is employed to converge the series in (14) is chosen to satisfy seven figure of accuracy.

Figs. 1 and 2 explain the complementary receiver characteristics (CROC) curve, i.e., $\bar{P}_{md}(\lambda) = 1 - \bar{P}_d(\lambda)$ versus $P_f(\lambda)$ over $\kappa - \mu$ shadowed and Fisher-Snedecor $\mathcal{F}$ fading scenarios, respectively, for different scenarios and $u = 2$. Figs. 3 and 4 show the complementary AUC, 1-$\bar{A}$ versus $\bar{\gamma}$ over $\kappa - \mu$ shadowed and Fisher-Snedecor $\mathcal{F}$ fading channels, respectively, for different scenarios and $u = 2$.

Figs. 5 and 6 demonstrate the effective rate versus the average SNR, $\bar{\gamma}$, over $\kappa - \mu$ shadowed and Fisher-Snedecor $\mathcal{F}$ fading scenarios, respectively, for different scenarios and $A = 1$.

In all figures, when $m$ in $\kappa - \mu$ shadowed and $m_s$ Fisher-Snedecor $\mathcal{F}$ is high, the shadowing impact is reduced. This would lead to improve the performance. Moreover, when $\kappa$ or/and $\mu$ increase, the detection capability of CRU and the effective rate become better. This refers to high power of dominant component in comparison with the scattered parts and the large number of clusters that arrives at the receiver, respectively.

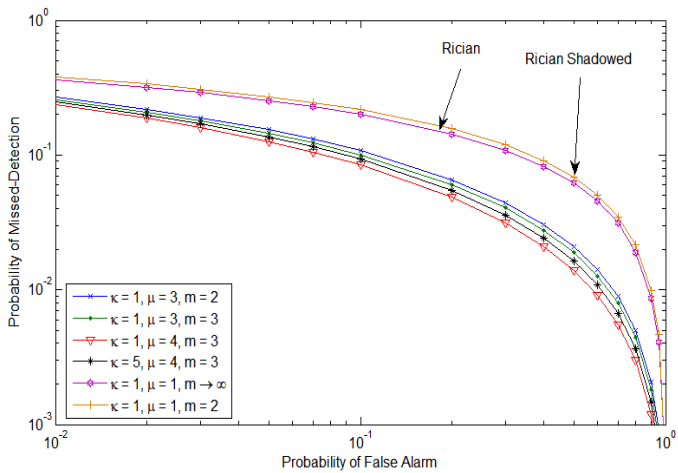

Fig. 1. The CROC in $\kappa - \mu$ shadowed fading scenarios for various $\kappa, \mu, m, u = 2$, and $\bar{\gamma} = 10$ dB.

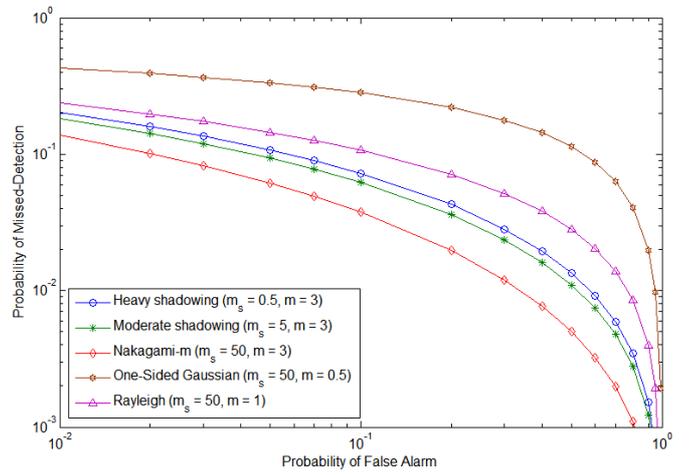

Fig. 2. The CROC over Fisher-Snedecor $\mathcal{F}$ fading scenarios for various $m$, $m_s$, $u = 2$, and $\bar{\gamma} = 0$ dB.

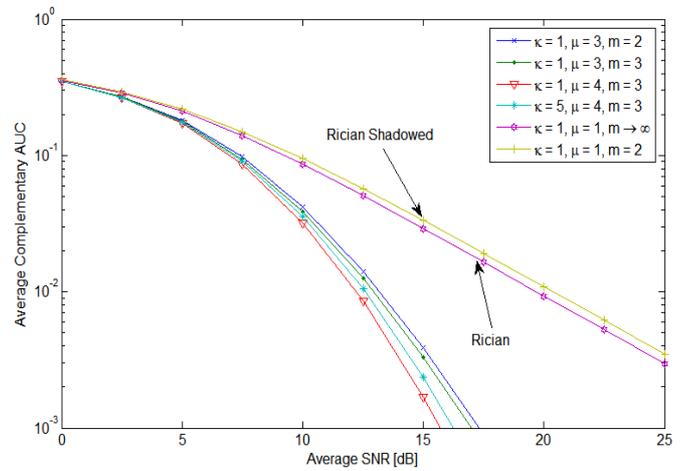

Fig. 3. The average complementary AUC in $\kappa - \mu$ shadowed fading scenarios for various $\kappa, \mu, m$ and $u = 2$.

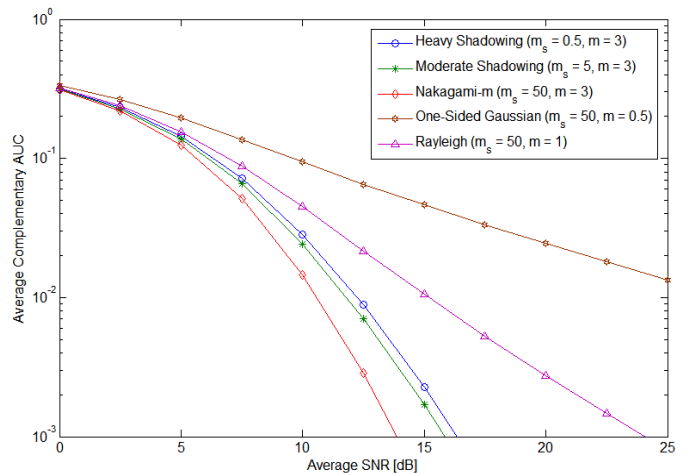

Fig. 4. The average complementary AUC over Fisher-Snedecor $\mathcal{F}$ fading scenarios for various $m$, $m_s$, and $u = 2$.

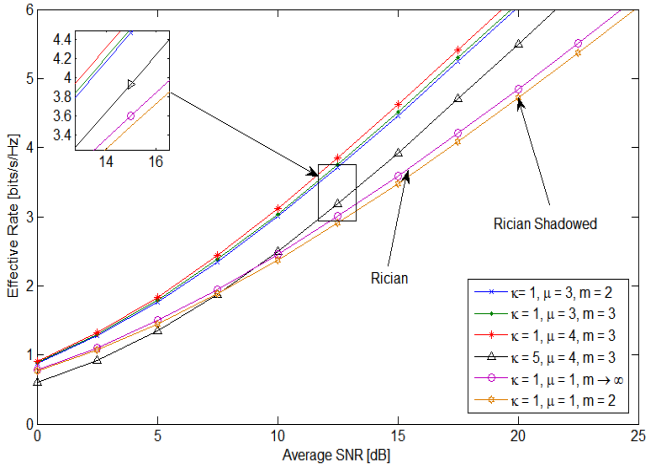

Fig. 5. The effective rate in $\kappa - \mu$ shadowed fading scenarios for various $\kappa$, $\mu$, $m$, and $A = 1$.

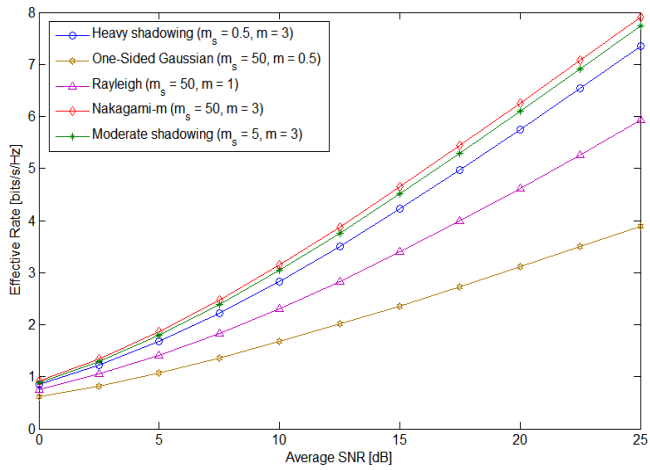

Fig. 6. The effective rate over Fisher-Snedecor $\mathcal{F}$ fading scenarios for various $m$, $m_s$, and $A = 1$.

## VII. CONCLUSIONS

The performance of CR systems over $\kappa - \mu$ shadowed and Fisher-Snedecor $\mathcal{F}$, is analyzed in this paper. A simple closed-form exact PDF of the SNR over $\kappa - \mu$ shadowed fading condition is derived first via supposing both $\mu$ and shadowing severity index are integer numbers. This PDF is then used to analyze the behavior of ED and effective rate of CDU. In the second part, the Fisher-Snedecor $\mathcal{F}$ fading condition which is recently proposed as a generalized composite distribution is employed. From the results, it can be noted that the performance becomes better when any of the fading parameters increase. Moreover, our derived results can provide a good analysis about the behavior of energy detection as well as the effective rate at different scenarios of shadowing impacts.

## REFERENCES

[1] M. K. Simon, and M.-S. Alouini, *Digital Communication over Fading Channels*, 2nd ed. New York: Wiley, 2005.

[2] A. Abdi and M. Kaveh, "$K$ distribution: an appropriate substitute for Rayleigh-lognormal distribution in fading-shadowing wireless channels," *Electron. Lett.*, vol. 34, no. 9, pp. 851-852, Apr. 1998.

[3] P. M. Shankar, "Error rates in generalized shadowed fading channels," *Wireless Personal Commun.*, vol. 28, no. 3, pp. 233-238, 2004.

[4] A. Laourine *et al.*, "On the performance analysis of composite multipath/shadowing channels using the $\mathcal{G}$-distribution," *IEEE Trans. Commun.*, vol. 57, no. 4, pp. 1162-1170, Apr. 2009.

[5] J. F. Paris, "Statistical characterization of $\kappa - \mu$ shadowed fading channels," *IEEE Trans. Veh. Technol.*, vol. 63, no. 2, pp. 518-526.

[6] H. Al-Hmood, and H. S. Al-Raweshidy, "Unified modeling of composite $\kappa - \mu$/gamma, $\eta - \mu$/gamma, and $\alpha - \mu$/gamma fading channels using a mixture gamma distribution with applications to energy detection," *IEEE Ant. and Wireless Propag. Lett.*, vol. 16, no. , pp. 104-108, 2017.

[7] H. Al-Hmood and H. S. Al-Raweshidy, "On the sum and the maximum of non-identically distributed composite $\eta - \mu$/gamma variates using a mixture gamma distribution with applications to diversity receivers", *IEEE Trans. Veh. Technol.*, vol. 65, no. 12, pp. 10048-10052, 2016.

[8] H. Al-Hmood, "A mixture gamma distribution based performance analysis of switch and stay combining scheme over $\alpha - \kappa - \mu$ shadowed fading channels," in Proc. *IEEE Annual Conf. on New Trends in Info. & Commun. Technol. Applications (NTICT'2017)*, 7-9 March 2017, pp. 292-297.

[9] M. Aloqlah, "Performance analysis of energy detection-based spectrum sensing in $\kappa - \mu$ shadowed fading," *Elect. Lett.*, vol. 50, no. 25, pp. 1944-1946, 2014.

[10] M. S. Aloqlah, I. E. Atawi, and M. F. Al-Mistarihi, "Further performance results for energy detector operating over $\kappa - \mu$ shadowed fading," *Proc. IEEE Pers. Ind. and Mob. Rad. Commun. (PIMRC)*, China, Aug.-Sept. 2015, pp. 668-671.

[11] G. Chandrasekaran and S. Kalyani, "Performance analysis of cooperative spectrum sensing over $\kappa - \mu$ shadowed fading," *IEEE Wireless Commun. Lett.*, vol. 4, no. 5, pp. 553-556, 2015.

[12] H. Al-Hmood, and H. S. Al-Raweshidy, "Analysis of energy detection with diversity receivers over non-identically distributed $\kappa - \mu$ shadowed fading channels," *Elect. Lett.*, vol. 53, no. 2, pp. 83-85, Jan. 2017.

[13] J. Zhang, L. Dai, W. H. Gerstacker, and Z. Wang, "Effective capacity of communication systems over $\kappa - \mu$ shadowed fading channels," *Elect. Lett.*, vol. 51, no. 19, pp. 1540-1542, 2015.

[14] H. Al-Hmood, and H. S. Al-Raweshidy, "Unified approaches based effective capacity analysis over composite $\alpha - \eta - \mu$/gamma fading channels," *Elect. Lett.*, vol. 54, no. 13, pp. 852-853, 2018.

[15] F. J. Lopez-Martinez, J. F. Paris, and J. M. Romero-Jerez, "The $\kappa - \mu$ shadowed fading model with integer fading parameters," *IEEE Trans. Veh. Technol.*, vol. 66, no. 9, pp. 7653-7662, 2017.

[16] S. K. Yoo, *et. al.*, "The Fisher-Snedecor $\mathcal{F}$ distribution: A simple and accurate composite fading model," *IEEE Commun. Lett.*, vol. 21, no. 7, pp. 1661-1664, 2017.

[17] I. S. Gradshteyn and I. M. Ryzhik, *Table of Integrals, Series and Products*, 7th ed. Academic Press Inc., 2007.

[18] F. F. Digham, M. S. Alouni, and M. K. Simon, "On the energy detection of unknown signals over fading channels," *IEEE Trans. Commun.*, vol. 55, no. 1, pp. 21-24, 2007.

[19] P. C. Sofotasios, M. Valkama, Y. A. Brychkov, T. A. Tsiftsis, S. Freear, and G. K. Karagiannidis,"Analytic solutions to a Marcum $Q-$function-based integral and application in energy detection," *in Proc. IEEE CROWNCOM*, Oulu, Finland, June 2014, pp. 260-265.

[20] H. Al-Hmood, "Performance analysis of energy detector over generalized wireless channels in cognitive radio," PhD Thesis, Brunel University London, 2015.